\documentclass[aps,twocolumn,showpacs,superscriptaddress]{revtex4}
\usepackage{amsmath}
\usepackage{amsfonts}
\usepackage{amssymb} 
\usepackage[dvips]{graphicx}
\usepackage{graphicx} 

\renewcommand{\textrm}{\rm} 

\newcommand{\ul}{\underline}
\newcommand{\td}{\tilde}
\newcommand{\Tr}{\mathop{\mathrm{Tr}}}
\newcommand{\Sign}{\mathop{\mathrm{Sign}}}

\newcommand{\vecb}[1]{{\bf{#1}}}
\newcommand{\upd}{{\mathrm{d}}}
\newcommand{\dvechat}{\hat{\vecb d}}
\newcommand{\nvechat}{\hat{\vecb n}}

\newcommand{\kvechat}{{\hat{\vecb k}}}

\newcommand{\xvechat}{\hat{\vecb x}}

\newcommand{\zvechat}{\hat{\vecb z}}

\newcommand{\fvec}{\vecb f}
\newcommand{\gvec}{\vecb g}

\newcommand{\vvec}{\vecb v}

\newcommand{\Rvec}{\vecb R}
\newcommand{\zerovec}{\vecb 0}

\newcommand{\nuvec}{\boldsymbol{\nu}}

\newcommand{\sigmavec}{\boldsymbol{\sigma}}
\newcommand{\nablavec}{\boldsymbol{\nabla}}
\newcommand{\Deltavec}{\boldsymbol{\Delta}}
\newcommand{\gammavec}{\boldsymbol{\gamma}}

\newcommand{\omegaJ}{\omega_{J}}

\newcommand{\TJ}{T_{J}}

\newcommand{\IS}{I^{\mathrm{S}}}
\newcommand{\IC}{I^{\mathrm{C}}}

\newcommand{\vF}{v_{\mathrm{F}}}
\newcommand{\vFvec}{\vvec_{\mathrm{F}}}

\newcommand{\Tc}{T_{\mathrm{c}}}
\newcommand{\kB}{k_{\mathrm{B}}}

\newcommand{\iu}{{\mathrm{i}}}

\newcommand{\im}{\mathop{\textrm{Im}}}
\newcommand{\re}{\mathop{\textrm{Re}}}

\newcommand{\fold}{\circ}
\newcommand{\kel}{\check}
\newcommand{\nam}{\hat}
\newcommand{\spin}{}
\newcommand{\lside}{{\textrm{l}}}
\newcommand{\rside}{{\textrm{r}}}

\begin{document}

\title{Multiple Andreev Reflections
in Weak Links of
Superfluid $^3$He-B}

\author{J. K. Viljas}
\affiliation{Low Temperature Laboratory, Helsinki University of
Technology, P.O.Box 2200, FIN-02015 HUT, Finland } 

\date{\today}

\begin{abstract}
We calculate the current-pressure characteristics of a 
ballistic pinhole aperture between two volumes of B-phase
superfluid $^3$He.
The most important mechanism contributing to dissipative currents in
weak links of this type is
the process of multiple Andreev reflections.
At low biases this process is significantly affected by
relaxation due to inelastic
quasiparticle-quasiparticle collisions. 
In the numerical calculations, suppression of the superfluid order
parameter 
at surfaces is taken into account self-consistently. When this effect
is neglected, the theory may be developed analytically like in the 
case of $s$-wave superconductors. A comparison with experimental 
results is presented.

\end{abstract}

 
\maketitle

\section{Introduction}

Liquid $^3$He is a strongly interacting system
of fermionic atoms with nuclear spin 1/2.
Its superfluid state below the critical temperature 
$\Tc\approx 1$ mK  is characterized by the creation of a condensate
where the atoms form Cooper pairs \cite{vw}. 
This is similar to what happens for electrons in superconducting
metals and, although $^3$He atoms are electrically neutral,
many analogues exist between the transport properties of the two
physical systems. 
For example, in both systems so-called Andreev
reflection can exist, where quasiparticles are converted between
particle-like and hole-like branches of the excitation spectrum
by the pairing potential \cite{KurkijarviRainer}.
However, instead of the singlet $s$-wave state of conventional
superconductors, the pairing state in superfluid $^3$He exhibits 
spin-triplet $p$-wave symmetry. This means that the condensate has
internal degrees of freedom, resulting in a complicated
structure for the order parameter, and in the existence of multiple
superfluid phases. There is also no crystal potential to 
impose symmetry restrictions, 
as in the case of unconventionally
paired ($d$-wave) superconductors.
As a result, many of the analogous phenomena occur in a more 
complicated form in $^3$He than anywhere else. 
In this paper we study the properties
of pressure-biased weak links in superfluid $^3$He.
The weak links consist of small apertures in a wall between two
volumes of the liquid \cite{DavisPackardRMP}, and, as such, are
analogous to ballistic point contacts between superconducting metals.
The theory of superconductor point contacts is well
developed, and thus most of the basic ideas 
may simply be inferred from existing results
\cite{KulikOmelyanchuk,Zaitsev,GunsenheimerZaikin,AverinBardas,AverinImam,Octavio}.

Most importantly, due to Andreev reflection, there
are bound quasiparticle states localized at the weak link, 
whose energies are below the bulk gap $\Delta$ \cite{Viljas4}.
These sub-gap states 
are responsible for 
carrying the phase-dependent supercurrents, 
\emph{i.e.},
the Josephson effects \cite{DavisPackardRMP,Viljas2}. 
When the contact is biased by a chemical potential
difference $U$, the supercurrents oscillate
at the Josephson frequency $2U/\hbar$.
Under such a bias, also dissipative dc currents will be generated.
The most obvious source of such currents is due to
thermally excited quasiparticles, but the resulting current 
is very small at low temperatures.
However, in the case of
a point contact, the Cooper pairs themselves may
participate in the flow of a dissipative current.
This is because in transmitting a
pair between the two condensates, energy can be conserved by
transferring the excess energy $2U$
to the bound-state quasiparticles. 
As a result, large dc currents can flow with arbitrarily small biases 
$U\ll\Delta$ also at low temperatures.
The underlying process by which this is accomplished 
is known as multiple Andreev reflections (MAR) \cite{Octavio}. 
In this process, the
bound quasiparticles are Andreev-reflected several times from the 
surrounding pairing potentials under the influence of the bias $U$. 
After two successive retroreflections, a quasiparticle has gained the
energy $2U$, and this corresponds to the dissipative transmission of 
one Cooper pair. 
This coherent process is repeated until the bound quasiparticles 
escape to energies above
$\Delta$, or are relaxed by inelastic scattering. 
The maximum number $n_{}$ of sub-gap reflections is given 
by $n_{}U=2\Delta$.
In superconductor weak links with non-perfect transparency, MAR
can give rise to a highly pronounced ``subharmonic gap structure'' (SGS), 
where the differential conductance is peaked at the biases
$U=2\Delta/n$, with $n=1,2,3,\ldots$ \cite{Octavio}.
On the other hand, in the limit of very low transparency, a tunnel junction is formed,
where the sub-gap states and hence MAR are completely suppressed.

In the case of superfluid $^3$He the SGS is not likely to be
observable in practice. There are two reasons for this.
First, the practically achievable
weak link diameters are quite large:
generally on the order of the zero-temperature
coherence length ($\xi_0\approx 10\ldots70$ nm), and certainly much 
larger than the Fermi wavelength ($\lambda_{\textrm{F}}\approx 0.8$ nm).
Since liquid $^3$He is naturally free of impurities,
the quasiparticles simply follow classical ballistics 
through the aperture. Some non-transparency is introduced by 
scattering at the
walls inside a finite-length aperture, but this scattering
is diffusive and its principal effect is to reduce the net 
currents \cite{Viljas2}.
Second, also for practical reasons, the biases $U$ in weak links of
$^3$He are always restricted to the limit $U\ll\Delta$
\cite{Steinhauer,Simmonds}, while the SGS
occurs on the scale of $\Delta$. In fact, in most experiments
$U$ is even much smaller than the quasiparticle relaxation strength 
$\hbar\Gamma$ due to inelastic scattering, which by itself satisfies 
$\hbar\Gamma\ll\Delta$.
The limit of a low-transparency point contact between 
triplet-paired condensates was recently studied \cite{bolech},
but, as explained above, such results are not likely to be important
for interpreting 
experiments in superfluid $^3$He. For intermediate transparencies,
effects similar to those of Ref.\ \onlinecite{kopnin} may be expected.

In this paper we consider the limit of a short
point contact with perfect transparency, the so-called
``pinhole''. 
Furthermore, we concentrate on studying the bias region $U\ll\Delta$, 
and consider only the B phase of superfluid $^3$He explicitly.
However, some of the general results may just as well be applied for 
the A phase, or any other triplet or singlet pairing state, and for
any value of a constant bias $U$. 
Even though the SGS in the dc current 
cannot be resolved with our assumptions, there are 
other details introduced by the complicated structure of the 
order parameter in $^3$He, and its modification due to surface 
scattering.
The equilibrium limit $U=0$ for a $^3$He-B pinhole was studied 
in Ref.\ \onlinecite{Viljas2} in detail, and this paper represents 
an generalization of that calculation to finite biases.
Parts of our results have already been published \cite{Viljas4} 
and we review them here in order to obtain a self-contained presentation. 
In addition, we present some new analytical results and
a more thorough numerical analysis of the dc current and 
supercurrent amplitudes as a function of the bias pressure.
Some aspects related to the so-called ``anisotextural'' effects \cite{Viljas2} 
are covered in more detail elsewhere \cite{Viljas5}.

In Sec.\ \ref{s.qc} we start with some basic issues 
of the quasiclassical theory, and in Sec.\ \ref{s.general}
the pinhole model and the general current formulas are introduced.
Section \ref{s.nosupp} presents the analytical results obtained
when surface pair-breaking is neglected. In this case many
limiting cases are studied, and we also briefly discuss the 
connection of the quasiclassical model to the anisotextural
effects \cite{Viljas2}.
In Sec.\ \ref{s.num} we present our numerical 
results for the current amplitudes
and the sub-gap bound states in the presence of the pair-breaking
effects, and the computational methods are briefly explained.
A comparison of the results to experimental data is provided in 
Sec.\ \ref{s.comparison}, and the agreement is found to be good.
Section \ref{s.disc} concludes with some discussion of future
directions. Finally, details related to the self-consistent 
computation of the order parameters and some mathematical results
are gathered in the Appendices.

\section{Quasiclassical framework} \label{s.qc}

Our analysis is based on the nonequilibrium formulation of 
quasiclassical theory,
which has been throughly reviewed in Ref.\ \onlinecite{SereneRainer}. 
We start by considering some basic points of the formalism here, since
it is of essential importance to the ensuing discussion. 
The central quantity is the Keldysh-space propagator, which 
has the form
\begin{equation} \label{e.prop}
\kel g = \left[\begin{matrix}
\nam g^R & \nam g^K \\ \nam 0 & \nam g^A
\end{matrix}\right], 
\quad \kel g \fold \kel g = -\pi^2\kel 1, 
\end{equation}
where $\nam g^{R,A,K}(\kvechat,\Rvec;\epsilon,t)$ 
are $4\times 4$ Nambu matrices,
and ``$\fold$'' denotes the quasiclassical folding product
\cite{SereneRainer}
--- see Appendix \ref{s.app1}. 
Here $\kvechat$ parametrizes positions on the spherical Fermi surface
of $^3$He, $\Rvec$ is a spatial coordinate, $\epsilon$ the
quasiparticle energy, and $t$ is time. The Nambu matrices have the
structure
\begin{equation} \label{e.gspincomp}
\nam g^{R,A} = \left[\begin{matrix}
\spin g_{}^{R,A} & \spin f_{}^{R,A} \\
\spin{\td f}_{}^{R,A} & \spin{\td g}_{}^{R,A}
\end{matrix}\right], \quad
\nam g^{K} = \left[\begin{matrix}
\spin g_{}^{K} & \spin f_{}^{K} \\
-\spin{\td f}_{}^{K} & -\spin{\td g}_{}^{K}
\end{matrix}\right]
\end{equation}
where the diagonal components $\spin{g}_{}^{R,A,K}$ and 
off-diagonal components $\spin{f}_{}^{R,A,K}$ are $2\times 2$ spin
matrices, and
the conjugation operation ``$~\td~~$'' is defined as
$\spin{\td q}(\kvechat,\epsilon)=\spin{\td q}(-\kvechat,-\epsilon)^*$.
In order to automatically satisfy the normalization condition in 
Eq.\ (\ref{e.prop}), it is
convenient to parametrize the propagator as follows 
\cite{SchopohlMaki,NagatoNagaiHara,Eschrig}:
\begin{equation}
\nam g^{R,A}=\mp\iu\pi\nam N^{R,A}\fold
\left[\begin{matrix}
\spin 1+\spin{\gamma}^{R,A}\fold\spin{\td\gamma}^{R,A} & 2\spin{\gamma}^{R,A} \\
-2\spin{\td \gamma}^{R,A} & -\spin 1-\spin{\td\gamma}^{R,A}\fold\spin{\gamma}^{R,A}
\end{matrix}\right]
\end{equation}
and
\begin{equation} \label{e.keldysh}
\begin{split}
\nam g^K&=-2\pi\iu\nam N^R \\
&\fold\left[\begin{matrix}
(\spin{x}^K-\spin{\gamma}^R\fold\spin{\td x}^K\fold\spin{\td\gamma}^A) &
-(\spin{\gamma}^R\fold\spin{\td x}^K-\spin{x}^K\fold\spin\gamma)^A \\
-(\spin{\td\gamma}^R\fold\spin x^K-\spin{\td x}^K\fold\spin{\td\gamma}^A)
&
(\spin{\td x}^K-\spin{\td\gamma^R}\fold\spin{x}^K\fold\spin{\gamma^A}) &
\end{matrix}\right] \\ 
&\qquad\fold\nam N^A,
\end{split}
\end{equation}
where 
\begin{equation}
\nam N^{R,A}=\left[\begin{matrix}
(\spin 1-\spin{\gamma}^{R,A}\fold\spin{\td\gamma}^{R,A})^{-1} & 0 \\
0 & (\spin 1-\spin{\td\gamma}^{R,A}\fold\spin{\gamma}^{R,A})^{-1}
\end{matrix}\right].
\end{equation}
Here the spin matrices $\spin \gamma^{R,A}(\kvechat,\Rvec;\epsilon,t)$ are called
coherence functions, and they may often be interpreted as
Andreev-reflection amplitudes. Since they fully parametrize 
$\nam g^{R,A}$, they completely determine the
density of quasiparticle states of the system. The spin matrix 
$\spin x^K(\kvechat,\Rvec;\epsilon,t)$, on the other hand, 
is a distribution function describing the occupation of these states.
All expectation values of one-body observables may be
computed from the Keldysh component $\nam g^K$, which includes 
information on the states as well as their occupation.
The coherence functions satisfy the symmetry
$\spin{\td\gamma}^{R}=(\spin{\td\gamma}^A)^\dagger$ and are related to
the spin components of the propagator by
$\spin\gamma^R=-(\iu\pi-\spin g_{}^R)^{-1}\fold\spin f_{}^R$.
The distribution function is Hermitian: $(\spin x^K)^\dagger=\spin x^K$.

The function $\spin x^K$ is not the only way to 
introduce a distribution function.
A more common definition is given by writing
\begin{equation}
\nam g^K=\nam g^R\fold\nam h - \nam h\fold g^A,
\end{equation}
which satisfies the normalization condition for any $\nam h$.
However, any physical $\nam g^K$ may be parametrized by choosing
$\nam h$ diagonal
\begin{equation}
\nam h=\left[\begin{matrix}
\spin h_1 & 0 \\
0 & -\spin {\td h}_1 
\end{matrix}
\right],
\end{equation}
with the spin matrix $\spin h_1(\kvechat,\Rvec;\epsilon,t)$ as the new
distribution function. 
Also $\spin h_1$ is Hermitian, $(\spin h_1)^\dagger=\spin h_1$,
and it is connected to $\spin x^K$ by the relation
\begin{equation} \label{e.qpdist}
\spin h_1 = \sum_{n=0}^\infty
(\spin{\gamma}^R\fold\spin{\td\gamma}^R)^n\fold
[\spin{x}^K-\spin{\gamma}^R\fold\spin{\td x}^K\fold\spin{\td\gamma}^A]\fold
(\spin{\gamma}^A\fold\spin{\td\gamma}^A)^n.
\end{equation}
The functions $\spin{x}^K$ and $\spin{\td x}^K$ have the interpretations of
distribution functions for ``particle-like'' and ``hole-like''
excitations, while $\spin h_1$ includes contributions from the coherent
Andreev reflections between the two types. 
In equilibrium $\spin h_1$ reduces to the function
$h(\epsilon)=\tanh(\beta\epsilon/2)=1-2f(\epsilon)$, where $\beta=1/\kB T$, 
$T$ is the temperature, and $\kB$ is Boltzmann's constant, and
$f(\epsilon)$ is the Fermi distribution.
In comparison, $\spin x^K$ takes the form
$\spin x^K(\epsilon)=h(\epsilon)(\spin 1 - \spin\gamma^R\spin{\td\gamma}^A)$.
Thus $\spin h_1$ has a more direct interpretation as a ``quasiparticle''
distribution function in, for example, the Andreev bound
states inside the weak link or a vortex core. 

The propagator $\kel g$ satisfies a transport-like equation of motion, 
which depends on self-consistently computed self-energies. The latter have 
a similar Keldysh-space and Nambu-space structure as 
$\kel g$ in Eqs.\ (\ref{e.prop}) 
and (\ref{e.gspincomp})\cite{SereneRainer,Eschrig}. 
The equation for $\kel g$ may be rewritten as a 
Riccati-type transport equation 
for the coherence functions $\spin{\gamma}^{R,A}$, and 
a kinetic equation for $\spin x^K$ \cite{Eschrig}.
In particular, the equation for $\spin\gamma^R$ is
\begin{equation} \label{e.riccati}
\begin{split}
\iu\hbar\vFvec\cdot\nablavec \spin{\gamma}^R
=&-2\epsilon\spin{\gamma}^R
-\spin{\Delta}^{R} \\
&+\spin{\gamma}^R\fold\spin{\td \Delta}^R\fold\spin{\gamma}^R 
+\spin{\Sigma}^R\fold\spin{\gamma}^R-\spin{\gamma}^R\fold\spin{\td\Sigma}^R
\end{split}
\end{equation}
where the spin matrices $\spin\Sigma^R$ and $\spin\Delta^R$ are the Nambu-space diagonal
and off-diagonal self energies, respectively, and $\vFvec=\vF\kvechat$
is the Fermi velocity. 
However, we only need these in equilibrium, where the kinetic
equation is always solved by
$\spin x^K=h(\spin 1-\spin\gamma^R\spin{\td\gamma}^A)$,
and the folding products in Eq.\ (\ref{e.riccati}) simplify to matrix products.
In the mean-field approximation $\spin\Sigma^{R,A}=\spin\Sigma^{\rm mf}$ and 
$\spin\Delta^{R,A}=\spin\Delta^{\rm mf}$, which are independent of $\epsilon$. 
Most importantly, the off-diagonal spin matrix $\spin\Delta^{\rm mf}$
determines the order parameter of the superfluid. 
In this paper the strong-coupling effects, 
\emph{i.e.}, inelastic quasiparticle-quasiparticle scattering, 
are only taken into account with a simple ``normal-state'' model 
$\spin\Sigma^{R,A}=\spin\Sigma^{\rm mf}\pm\iu\Gamma_1(\epsilon)$,
where $\Gamma_1(-\epsilon)=\Gamma_1(\epsilon)$. This
effectively adds an imaginary part to energies: 
$\epsilon\rightarrow\epsilon^{R,A}=\epsilon\pm\iu\Gamma_1(\epsilon)$.
Physically, the imaginary part describes a finite quasiparticle lifetime, which 
is important in the parameter ranges of $^3$He weak link
experiments \cite{Steinhauer,Simmonds}. Mathematically, it is
important for regularizing the divergences in the MAR process, 
which occur at low pressure biases \cite{GunsenheimerZaikin}.
In general, the collisional self-energy \cite{SereneRainer} 
gives strong-coupling corrections also to $\spin\Delta^{R,A}$ (and
hence a gap-dependent contribution to the lifetime \cite{AverinBardas}), 
but their proper calculation is too complicated for the purposes of this
paper.

\section{Pressure-biased pinhole} \label{s.general}

\begin{figure}[!t]
\includegraphics[width=0.99\linewidth]{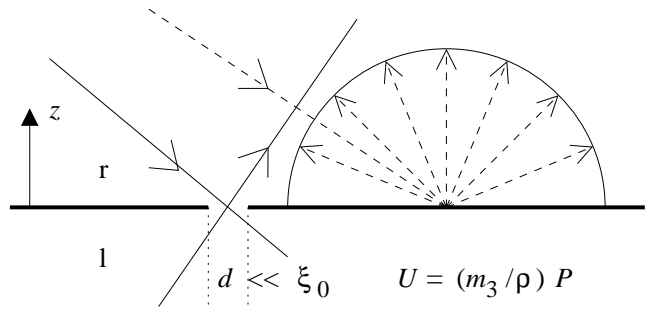}
\caption{Quasiparticles hitting the wall outside of the constriction
are scattered, which leads to a suppression of the superfluid state at
distances closer to the wall than $\xi_0$ (dashed trajectories). 
Only quasiparticles hitting the constriction directly are
ballistically transmitted and contribute to
the current (solid trajectories).}
\label{f.traj}
\end{figure}

We now apply the above formalism to describe a small constriction
of diameter $d$ and area $S=\pi(d/2)^2$
in wall between two volumes of superfluid $^3$He, when there is a
pressure difference $P$ between the two sides. 
As a result of thermomechanical effects, 
there may also exist a temperature difference. 
Although we initially allow for this possibility, we shall always 
assume the sides to be in good thermal
contact and thus at equal temperatures.
We use the so-called
``pinhole'' model, which is a direct generalization of that used
already in Ref.\ \onlinecite{KulikOmelyanchuk}
--- see Fig.\ \ref{f.traj}. 
Thus we assume that while
$d\gg\lambda_{\textrm{F}}$, it is still much smaller than
the zero-temperature coherence length
$\xi_0=\hbar\vF/2\pi\kB\Tc$. We also assume that the 
wall is negligibly thin in comparison with $d$, 
so that scattering inside the aperture need not be
considered. (A simple way relax the latter assumption was
considered in Ref.\ \onlinecite{Viljas2}.)
The convenience of this
model is that the calculation of the nonequilibrium current through
the constriction 
need not be done self-consistently, since all feedback effects 
away from the aperture may be neglected to lowest order in $d/\xi_0$.
It is enough the know the \emph{equilibrium} propagators (or coherence
functions) 
calculated close to the wall on its left ($\lside$) and right
($\rside$) sides in the absence of the constriction.
However, in anisotropically paired superfluid $^3$He the
computation of these propagators
still requires a one-dimensional self-consistent calculation, 
since the presence of the 
surface leads to pair-breaking effects
\cite{AmbegaokarDegennesRainer} and to the existence of surface-bound
quasiparticle states below the bulk gap.

In the following we assume that the order parameters and the
corresponding equilibrium coherence functions $\spin \gamma^{Ri}$ and
$\spin \gamma^{Ai}$
have already been calculated on both sides $i=\lside,\rside$ ---  
see Sec.\ \ref{s.num} and Appendix 
\ref{s.app1} for more details.
Thus the results of this section are still very general and applicable
to any type of pairing state.

Let us choose the coordinates such that the $z$ axis is perpendicular
to the wall and points from $\lside$ ($z<0$) to $\rside$ ($z>0$).
The current through the pinhole is given by \cite{SereneRainer}
\begin{equation}
\begin{split}
I(t)=&G_{\rm n} 
\int\frac{\upd\epsilon}{2\pi\iu} 
\left\langle \hat k_z  
{\Tr} \spin{C(\epsilon)}
\left[
\spin g_{}^K(\kvechat)-\spin
g_{}^K(-\kvechat)
\right]
\right\rangle_{\hat k_z>0}
\end{split}
\end{equation}
where $\spin g_{}^K(\kvechat;\epsilon,t)$ is the diagonal 
Nambu component of $\nam g^K$
inside
the pinhole, 
$\langle\cdots\rangle_{\hat k_z>0}=\int_{\hat k_z>0}
(\upd\Omega_{\kvechat}/4\pi)(\cdots)$,
and ${\Tr}$ is a spin-matrix trace. 
For particle current the spin matrix 
$\spin{C(\epsilon)}=\spin{1}$. For heat current 
$\spin{C(\epsilon)}=\epsilon$ and 
the spin current for spin projection along the axis $\alpha=x,y,z$
would be obtained by using the corresponding Pauli matrix 
$\spin{C}=\spin{\sigma}_\alpha$.
The unit $G_{\rm n} = \frac{1}{2}\vF N(0) S = M/\pi\hbar$
is the normal-state conductance, 
where $N(0)$ is the single-spin density of states in the normal state
\cite{Viljas2} and $M$ is the number of conducting transverse modes.

If we count energies from the chemical potential of the
side $\rside$, the pressure bias $P$ 
causes a shift in the $\lside$-side
chemical potential by $U=\mu^\lside-\mu^\rside$, where $U=(m_3/\rho)P$, $m_3$
being the mass of a $^3$He atom and $\rho$ the mass density of the liquid.
The phase difference $\phi=\varphi^\rside-\varphi^\lside$ between 
the $\lside$ and $\rside$ 
condensates then varies 
according to $\dot\phi=2U/\hbar$. Assuming $U$ to be constant, this is
solved by $\phi(t)=\omega_J t$ where the Josephson frequency
$\omegaJ=2U/\hbar$.
The constant bias $U$ gives $\spin\gamma^{R\lside}$ a simple time dependence,
$\spin\gamma^{R\lside}(\epsilon)\rightarrow $
$\gamma^{R\lside}(\epsilon,t)=e^{-\iu\phi(t)}\spin\gamma^{R\lside}(\epsilon)$,
while $\spin\gamma^{R\rside}$ remains time-independent.
This allows one to evaluate the folding products in $\spin g_{}^K$
[Eq.\ (\ref{e.keldysh})] analytically --- see Appendix \ref{s.app2}. 
First it is convenient to define 
\begin{equation} \label{e.genecurr}
I(t)=G_{\rm n}\left\langle \hat k_z I(\kvechat,t)
\right\rangle_{\hat k_z>0}.
\end{equation}
where $I(\kvechat,t)$ is
a ``channel-resolved'' current.
Since this
is periodic with the Josephson period 
$\TJ=2\pi/\omegaJ$, we expand
\begin{equation} \label{e.expansion}
I(\kvechat,t)=\sum_{m=-\infty}^{\infty}
I_m(\kvechat)e^{-\iu m\omegaJ t},
\end{equation}
where
$I_m(\kvechat)=I^*_{-m}(\kvechat)=\langle I(\kvechat,t)e^{\iu m\omegaJ
t}\rangle_{\TJ}$ and we defined the time average
$\langle\cdots\rangle_{T_J}={\TJ}^{-1}\int_0^{\TJ}(\cdots)\upd t$.
Evaluating the folding products one finds,
for $m\geq0$ and $\hat k_z>0$
\begin{equation} \label{e.intermediate}
\begin{split}
I_m(\kvechat)=&
-\sum_{n=0}^\infty \int\upd\epsilon  
\Tr \spin C(\epsilon)
\\
\bigg\{ &
\spin F_{\lside\rside}^{n+m,n}[\kvechat,\epsilon-(2n+m)U,U] \\
&-\spin F_{\rside\lside}^{n,n+m}[-\kvechat,\epsilon+(2n+m+1)U,-U]\bigg\}
\end{split}
\end{equation}
where 
\begin{equation} \label{e.fdef}
\begin{split}
\spin F_{ij}^{k,l}(\kvechat,\epsilon,U)=&
\spin P^k_{ij}(\epsilon,U) 
[\spin x_{}^{Ki}(\epsilon)- 
\spin\gamma_{}^{Ri}(\epsilon) \\
&\times\td{\spin x}_{}^{Kj}(\epsilon-U)
\td{\spin\gamma}_{}^{Ai}(\epsilon)]
[\spin P^l_{ij}(\epsilon,U)]^\dagger
\end{split}
\end{equation}
and
\begin{equation} \label{e.pdef}
\begin{split}
\spin P^k_{ij}(\epsilon,U) = 
\prod_{p=1}^{k}
\spin\gamma^{Ri}_{}(\epsilon+2pU)
\td{\spin\gamma}^{Rj}_{}(\epsilon+(2p-1)U).
\end{split}
\end{equation}
The distribution function is
$\spin x^{Ki}=h^i(\epsilon)$$[\spin
1-\spin\gamma^{Ri}\td{\spin\gamma}^{Ai}]$, and again
$i=\lside,\rside$. 
The $\kvechat$ dependences have been dropped for clarity.

If $\spin C(\epsilon)$ is assumed to be energy-independent, then 
Eq.\ (\ref{e.intermediate}) can be simplified by changing integration
variables.
In the following we also assume the two sides to be at equal
temperatures, such that
$h^{\lside,\rside}(\epsilon)=h(\epsilon)=\tanh(\beta\epsilon/2)$.
When the normal-state contribution proportional to
$\int\upd\epsilon[h(\epsilon+U)-h(\epsilon)]=2U$ is separated, 
one finds  
\begin{equation} \label{e.ccomp}
\begin{split}
I_m(\kvechat)=&
\Tr\spin C \bigg\{ 
2U\delta_{m0}-
\sum_{n=0}^\infty
\int\upd\epsilon  
\\
&[\spin F_{\lside\rside}^{n+m,n}(\kvechat,\epsilon,U)- 
\spin F_{\rside\lside}^{n,n+m}(-\kvechat,\epsilon,-U)]\bigg\}
\end{split}
\end{equation}
Identifying the coherence function $\spin\gamma^{R}$ 
as an Andreev-reflection amplitude, 
Eq.\ (\ref{e.ccomp}) has a clear interpretation as describing the MAR
process, with the index $n$ running over the 
number of successive reflections.
The present results have been derived by assuming $U$ to be constant.
However, it may be shown that even when $U$ varies in time, 
corrections to the results are small at least if
$\dot U\ll \hbar\Gamma_1^2$. When $U\ll\hbar\Gamma_1$ and 
$\dot U\lesssim \hbar\omegaJ^2$ this should be well satisfied.

The current
may also be Fourier-expanded as
\begin{equation} \label{e.fullc}
\begin{split}
I(U,t)=&I_{0}(U)+\sum_{m=1}^\infty[
\IS_{m}(U)\sin(m\omegaJ t) \\
&+
\IC_{m}(U)\cos(m\omegaJ t)
]
\end{split}
\end{equation}
where the coefficients $I_0,\IS_m,\IC_m$ are real-valued.
They
are connected to the complex amplitudes of Eq.\ (\ref{e.expansion}) by 
\begin{equation}
\begin{split}
I_0&=G_{\rm n}\langle \hat k_zI_0(\kvechat)\rangle_{\hat k_z>0}, \\
\IS_m&=2G_{\rm n}\im\langle \hat k_zI_m(\kvechat)\rangle_{\hat k_z>0},\\
\IC_m&=2G_{\rm n}\re\langle \hat k_zI_m(\kvechat)\rangle_{\hat k_z>0}.
\end{split}
\end{equation}
From Eqs.\ (\ref{e.keldysh}) and (\ref{e.qpdist}) we also note that
$\langle \spin g^K\rangle_{\TJ}=-2\pi\iu \spin h_1$ 
and thus the dc component is given by
\begin{equation} \label{e.dcc}
I_0=-G_{\rm n}\int\upd\epsilon\left\langle\hat k_z
\Tr C(\epsilon)
\left[\spin h_1(\kvechat)-\spin h_1(-\kvechat) \right]\right\rangle_{\hat k_z>0}.
\end{equation}
As seen in Eq.\ (\ref{e.ccomp}), it is convenient to separate the dc
current as
$I_{0}(U)=G_{\textrm{n}}U+I_{AR}(U)$.
Here $G_{\rm n}U$ is the normal-state part and 
and $I_{AR}(U)$ is due to MAR only \cite{GunsenheimerZaikin}. 
At high biases $I_{AR}$ saturates, and gives rise to the 
``excess current'' on top of $G_{\rm n}U$ (see below).
In what follows we shall be interested in calculating 
$I_0,\IS_m,\IC_m$ both analytically and numerically for the case of
superfluid $^3$He-B.
We only concentrate on analyzing the particle (or mass) current,
where $\spin C = \spin 1$.

\section{Results for the case with no gap suppression} \label{s.nosupp}

\subsection{General results} \label{s.genres}

For simplicity we shall first neglect the suppressing effect of the
solid wall on the order parameter.  This makes the problem formally
similar to  the $s$-wave case, and the results of this section are
rather straightforward generalizations of those of
Refs.\ \onlinecite{Zaitsev,GunsenheimerZaikin,AverinBardas,AverinImam}.
The B-phase order parameter \cite{vw} is thus assumed to be of the form 
$\spin\Delta^i(\kvechat,z)$
$\equiv\Deltavec^i(\kvechat)\cdot\spin\sigmavec\iu\spin\sigma_2$
for $i=\lside,\rside$, and $\spin\Sigma^{\rm mf}=0$.
In this case Eq.\ (\ref{e.riccati}) is easily solved \cite{Viljas4}.
The gap vectors for momentum direction $\kvechat$ are given by 
$\Deltavec^{\lside,\rside}(\kvechat)=$
$\Delta e^{\iu\varphi^{\lside,\rside}}\dvechat^{\lside,\rside}(\kvechat)$,
where $\dvechat^{\lside,\rside}(\kvechat)=R^{\lside,\rside}\kvechat$. Here
$R^{\lside,\rside}=R(\nvechat^{\lside,\rside},\theta_L)$ are rotation 
matrices, with
$\theta_L=\arccos(-1/4)$ the dipole-locked rotation angle \cite{vw},
and $\nvechat^{i}$ the rotation axis on side $i=\lside,\rside$.

If, for each $\kvechat$, we choose the spin quantization 
axis parallel to $\dvechat^\lside\times\dvechat^\rside$, 
the condensates may be 
divided into $\uparrow\uparrow$ and $\downarrow\downarrow$
parts, which behave much like two independent $s$-wave systems \cite{Yip}.
If we define $\phi_{\kvechat}^{\lside,\rside}$ as the azimuthal angles of
$\dvechat^{\lside,\rside}$ in the plane perpendicular to 
$\dvechat^\lside\times\dvechat^\rside$, which satisfy 
$\phi_{-\kvechat}^i=\phi_{\kvechat}^i+\pi$, then 
$\spin\Delta^{\lside,\rside}$ and $\spin\gamma^{\lside,\rside}$ 
are diagonal
\begin{equation} \label{e.diag1}
\spin\Delta^{\lside,\rside}=\Delta 
\left[\begin{matrix}
-e^{-\iu\phi_{\kvechat}^{\lside,\rside}} & 0 \\
0 & e^{\iu\phi_{\kvechat}^{\lside,\rside}}\end{matrix}\right], \quad
\end{equation}
\begin{equation} \label{e.diag2}
\spin\gamma^{R,A;\lside,\rside}=\gamma^{R,A} 
\left[\begin{matrix}
-e^{-\iu\phi_{\kvechat}^{\lside,\rside}} & 0 \\
0 & e^{\iu\phi_{\kvechat}^{\lside,\rside}}\end{matrix}\right].
\end{equation}
Here $\gamma^{R,A}=-\Delta/(\epsilon^{R,A}$
$\pm\iu\sqrt{\Delta^2-(\epsilon^{R,A})^2})$,
and $\epsilon^{R,A}=\epsilon\pm\iu\Gamma_1(\epsilon)$, where
$\Gamma_1$ is present to model inelastic scattering.
The phase differences of the two condensates 
over the contact are given by
$\phi_{\kvechat.\sigma}=\phi-\sigma\chi_{\kvechat}$,
where $\sigma=\pm 1$,
$\phi=\varphi^\rside-\varphi^\lside=\omegaJ t$, and
$\chi_{\kvechat}=\arccos(\dvechat^\lside\cdot\dvechat^\rside)$. 
Using these definitions, Eq.\ (\ref{e.ccomp})
simplifies to \cite{Viljas4}
\begin{equation} \label{e.final}
\begin{split}
I_m(\kvechat)=&\Tr\spin C\bigg\{ 
2U\delta_{m0} 
+2\left[
\begin{matrix}e^{\iu m\chi_{\kvechat}} & 0 \\
0 & e^{-\iu m\chi_{\kvechat}}
\end{matrix}\right] \\
&\times\mathcal{P}\int\upd\epsilon 
\tanh(\beta\epsilon/2) 
(1-|\gamma^R(\epsilon)|^2) \\
&\sum_{n=0}^\infty 
\prod_{q=1}^n|\gamma^R(\epsilon-qU)|^2
\prod_{p=l+1}^{n+2m}\gamma^R(\epsilon-pU) \bigg\}.
\end{split}
\end{equation}
The distribution function $\spin h_1$ is proportional to the 
unit matrix in spin space.
For $\hat k_z>0$ it is given by
\begin{equation}
\begin{split}
h_1^{>}(\epsilon) = &
h(\epsilon) + \sum_{n=0}^\infty \prod_{j=0}^n |\gamma^R(\epsilon-jU)|^2 \\
&\times[h(\epsilon-(n+1)U)-h(\epsilon-nU)]
\end{split}
\end{equation}
and for $\hat k_z<0$ by
\begin{equation}
\begin{split}
h_1^{<}(\epsilon) = &
h(\epsilon+U) + \sum_{n=0}^\infty \prod_{j=0}^n
|\gamma^R(\epsilon+(j+1)U)|^2 \\ 
&\times[h(\epsilon+(n+2)U)-h(\epsilon+(n+1)nU)].
\end{split}
\end{equation}
The dc component of the particle current [Eq.\ (\ref{e.dcc})]
may now be written
\begin{equation} \label{e.dch}
I_0 = -(G_{\textrm n}/2) \int\upd\epsilon [h_1^{>}(\epsilon)
-h_1^{<}(\epsilon)].
\end{equation}
We note that the result for $I_0$ is exactly the same as for an
$s$-wave superconductor \cite{AverinBardas,AverinImam}.
In particular, it is independent of the spin-orbit rotation matrices 
\cite{Viljas5}.

\subsection{Limiting cases} \label{s.limits}

In the small-bias (or adiabatic) limit $U\ll\Delta$ 
the variation of phase difference $\phi(t)=\omega_Jt$ is slow. In this
case one may describe the junction in
terms of the occupation of the Andreev bound states \cite{Viljas4}
\begin{equation} \label{e.states}
\epsilon_{\kvechat,\sigma}(\phi) = 
-\Sign(\hat k_z \sin(\phi_{\kvechat,\sigma}/2))
\Delta \cos (\phi_{\kvechat,\sigma}/2),
\end{equation}
which are obtained from the poles of $\nam g^R$ in equilibrium.
The Keldysh function may be now
approximated with the ``quasi-equilibrium'' form
$\nam g_{}^K=h_1^{\gtrless}(\nam g_{}^R-\nam g_{}^A)$ so that 
\begin{equation}
[\spin g^K_{}]_{\sigma\sigma}=
-4\pi^2\iu\Sign(\hat k_z)\frac{\upd\epsilon_{\kvechat,\sigma}(\phi)}
{\upd\phi}
\delta(\epsilon-\epsilon_{\kvechat,\sigma}(\phi))h_1^{\gtrless}(\epsilon).
\end{equation}
Defining the bound-state occupation probabilities
\begin{equation}
p_{\kvechat,\sigma}=
\left\{1-h_1^{\gtrless}[\epsilon_{\kvechat,\sigma}(\phi)]\right\}/2,
\quad \hat k_z\gtrless0
\end{equation}
the particle current may be written 
\begin{equation} \label{e.statecurrent}
I(t) = 4\pi G_{\rm n}\sum_{\sigma,\delta=\pm 1}
\left\langle\hat k_z 
\frac{\upd\epsilon_{\delta\kvechat,\sigma}(\phi)}{\upd\phi}
p_{\delta\kvechat,\sigma}
\right\rangle_{\hat k_z>0}.
\end{equation}
Neglecting Andreev reflections for $|\epsilon|>\Delta$, we may approximate
$\gamma^R\approx-e^{-\iu\vartheta(\epsilon)-\zeta(\epsilon)}\theta(\Delta-|\epsilon|)$
where 
$\vartheta(\epsilon)=\arccos(\epsilon/\Delta)$ and
$\zeta(\epsilon)=\Gamma_1(\epsilon)/\sqrt{\Delta^2-\epsilon^2}$
which is strictly valid only for $|\epsilon|\ll\Delta$.
Using these we may approximate $h_1^{>}$ as
\begin{equation}
\begin{split}
h_1^>(\epsilon)= &
h(\epsilon)-\theta(\Delta-|\epsilon|) 
\int_{-\Delta}^{\epsilon}\upd\epsilon'h'(\epsilon') \\
&\times
\exp[-\frac{2}{U}\int_{\epsilon'}^{\epsilon}\upd\epsilon''\zeta(\epsilon'')]
\end{split}
\end{equation}
and a similar expression exists for $h_1^<$.
Then it may be shown that the occupation probabilities satisfy the
kinetic equation
\begin{equation} \label{e.kineq}
\dot p_{\kvechat,\sigma}(t)=\Gamma(\epsilon)[f(\epsilon)-p_{\kvechat,\sigma}(t)],
\end{equation}
where $f=(1-h)/2$ is the Fermi distribution, 
$\epsilon=\epsilon_{\kvechat,\sigma}[\phi(t)]$,
and we defined the relaxation rate
$\Gamma=2\Gamma_1/\hbar$.
In a normal-state Fermi-liquid approximation
$\Gamma\sim(\pi\kB T)^2+\epsilon^2$, 
while in the superfluid state some corrections from
the existence of the gap may be expected \cite{AverinBardas}.
The initial conditions for this equation are mostly determined by 
the ``thermalization'' of the bound states when they hit the gap 
edges at $\epsilon=\pm\Delta$ \cite{AverinBardas,Viljas4}.
Thus if at $t=t_0$ we have
$\epsilon_{\kvechat,\sigma}[\phi(t_0)]=-\Delta$ (for $\hat k_z>0$), 
then the occupation is returned to equilibrium: 
$p_{\kvechat,\sigma}(t_0)=f(-\Delta)$.
In the limit $U\ll\hbar\Gamma$ Eq.\ (\ref{e.kineq}) may be solved to yield
\begin{equation}
\begin{split}
I(t)=&I_{\rm s}[\phi(t)] \\ 
&+G_{\rm n}U\sum_{\sigma,\delta}
\left\langle \hat k_z \frac{2\pi\beta}{\hbar\Gamma(\epsilon)}
\frac{[\upd\epsilon_{\delta\kvechat,\sigma}/\upd\phi]^2}{\cosh^2(\beta\epsilon/2)}
\right\rangle_{\hat k_z > 0}
\end{split}
\end{equation}
where $\epsilon=\epsilon_{\delta\kvechat,\sigma}[\phi(t)]$ and 
$I_{\rm s}(\phi)$ is the supercurrent of Ref.\ \onlinecite{Yip}.
Approximating $\Gamma(\epsilon)\approx\Gamma(0)\equiv\Gamma_0$,
its time average may be simplified to 
\begin{equation} \label{e.arc}
I_{0}(U)=(\Delta/\hbar\Gamma_0)g(T)G_{\rm n}U.
\end{equation}
where the temperature factor is given by
$g(T)=\int_{-1}^1\tanh(\beta\Delta x/2)(x/\sqrt{1-x^2})\upd x$.
This result is only correct to leading order in
$\hbar\Gamma_0/\Delta$, since we have neglected the corrections
from energies $|\epsilon|>\Delta$. This approximation is valid for
temperatures not too close to the critical temperature $T_c$.
Another exactly soluble limit of Eq.\ (\ref{e.kineq}) is that of 
$U\ll\Delta$ at zero temperature, if we additionally assume 
the Fermi-liquid form
$\Gamma(\epsilon)=c\epsilon^2$. 
In this case
\begin{equation} \label{e.zerot}
I_0(U) = 2G_{\rm n}\Delta\int_0^1\upd x \exp\left[-\frac{c\Delta^2}{2U}
(\arcsin x-x\sqrt{1-x^2})\right],
\end{equation}
which varies as $I_0(U)\sim U^{-1/3}$ when $U\rightarrow 0$, and not
linearly as Eq.\ (\ref{e.arc}) at higher temperatures.
This is plotted in Fig.\ \ref{f.i0} for experimentally feasible parameters.

For $U\sim\hbar\Gamma$ quasiparticles in the MAR cycle
begin to reach the gap without being scattered, and 
$I_{AR}$ begins to saturate. In the limit $\hbar\Gamma\ll U\ll \Delta$
the current is slowly varying, and on the order of
$2G_{\rm n}\Delta$. Finally, in the large-bias limit $U\gg\Delta$ 
one finds that $I_{AR}$ saturates with the asymptotic behavior 
\cite{Zaitsev,GunsenheimerZaikin}
\begin{equation} \label{e.exc}
I_{\rm exc}=\frac{8\Delta}{3} G_{\textrm{n}}\tanh(\beta U/2).
\end{equation}
This is known as an ``excess current'', since
the total dc current is then of the form
$I_0(U)=G_{\textrm{m}}U+I_{\rm exc}$, where the first term is the
normal-state value and the second one approaches a constant.


\subsection{Anisotextural effects}

Most of the above results have been derived by assuming the bias $U$
to be time-independent, and the spin-orbit textures to be fixed. 
Thus also $\chi_{\kvechat}$ must be constant in time.
However, there are situations where the textures may also oscillate 
resonantly with the Josephson frequency\cite{Viljas2},
and the theory may be generalized to take such ``anisotextural'' 
effects into account.
Consider in particular the limit $U\ll\hbar\Gamma_0$ and 
$T\approx \Tc$, which is realized in the experiments of 
Ref.\ \onlinecite{Simmonds}, for example.
Starting from Eqs.\ (\ref{e.statecurrent}) and (\ref{e.kineq})
we find the following expression
\begin{equation} \label{e.aniso}
\begin{split}
I(t)=&\pi G_{\rm n}\beta\Delta^2
\bigg\langle\hat k_z \Big\{
\cos\chi_\kvechat\sin\phi 
+ \frac{U}{\hbar\Gamma_0}\Big[1 \\
&-\cos\chi_\kvechat\cos\phi 
+\sin\chi_\kvechat
\frac{\hbar\dot\chi_\kvechat}{2U}\sin\phi\Big]\Big\}
\bigg\rangle_{\hat k_z>0}
\end{split}
\end{equation}
where $\phi=\omegaJ t$ and $\chi_\kvechat=\chi_\kvechat(t)$.
The first and third terms correspond to the 
$m=1$ terms in Eq.\ (\ref{e.fullc}), while the second term 
is the equivalent of Eq.\ (\ref{e.arc}).
The last term is new, and it is only present when 
$\chi_\kvechat$ is time-dependent.
Assuming now that
$\chi_\kvechat(t)$
oscillates $\TJ$-periodically, then we see that
all the $\phi$-dependent terms can also have finite time averages.
Actually, the averages of the last two terms in 
Eq.\ (\ref{e.aniso}) exactly cancel each other, but the average 
of the first term gives a dc current in addition to 
Eq.\ (\ref{e.arc}).
Thus, for $T\approx\Tc$ we may write
\begin{equation} \label{e.avg}
\begin{split}
I_{\rm dc, total}(U)\approx &
I_0(U)+
\langle \IS_1(U,t)\sin(\omegaJ t)\rangle_{\TJ} \\
\end{split}
\end{equation}
where $\IS_1$ has the $\TJ$-periodicity of
the angles $\chi_\kvechat(t)$.
The first term in Eq.\ (\ref{e.avg}) results from the MAR 
process, where energy is
dissipated directly to the quasiparticle system.
The second term corresponds to dissipation via the excitation of 
collective order-parameter modes, \emph{i.e.}, spin waves, which 
are driven by the oscillating Josephson spin currents \cite{Viljas2}.
The true magnitude of the resulting
dc current depends on the details of the process,
which is in general geometry-dependent.
Therefore, we shall not attempt to explore this issue any further here.
Note, however, that while the dc current $I_0(U)$ is independent 
of $R^{\lside,\rside}$, the second contribution may depend strongly on
them. In fact, for $U\ll\hbar\Gamma$
the coefficient $\IS_1$ is generally of the form
$\IS_1(t)=(2m_3/\hbar)[\alpha_1\psi_{zz}+\alpha_2(\psi_{xx}+\psi_{yy})](t)$,
where $\psi_{ij}=R_{\mu i}^{\lside}R_{\mu j}^{\rside}$. The
$U$-independent parameters $\alpha_1(T)$ and $\alpha_2(T)$ may be computed 
numerically even when gap suppression effects are accounted for 
\cite{Viljas2},
and expect Eq.\ (\ref{e.avg}) to remain valid also in this case.
Finally note that very similar dc current contributions
may arise from the ``Shapiro'' or ``Fiske'' type effects, 
where $\omegaJ$ coincides with
some resonance frequency of the cell instead of the order-parameter texture
\cite{BaronePaterno}.

\section{Gap suppression and numerics} \label{s.num}

\subsection{Calculation of the coherence functions} 
\label{s.calculation}

For a more realistic calculation, one must take into account the
suppression of the $p$-wave order
parameter 
$\spin\Delta_0=\Deltavec_0\cdot\spin\sigmavec\iu\spin\sigma_2$
close to solid walls. 
However, when this effect is included,
the calculation of the coherence functions may only be done
numerically, and the current must be computed from
Eq.\ (\ref{e.ccomp}).

We assume the surface to have translation symmetry in its plane,
and a rotation symmetry around its normal $\zvechat$. If we also 
neglect external magnetic fields and flow parallel to the surface, then 
the B-phase order parameter may be parameterized by the form
\begin{equation} \label{e.op}
\Deltavec_0(\kvechat,z)=
[\Delta_\parallel(z)\hat k_x,\Delta_\parallel(z)\hat k_y,
\Delta_\perp(z)\hat k_z],
\end{equation}
where $z=0$ is at the wall.
The real amplitudes 
$\Delta_\parallel(z)$ and $\Delta_\perp(z)$ 
both approach the value $\Delta$ as $|z|\rightarrow\infty$.
Close to the wall their behavior must be calculated using one of
the different models for the surface scattering, 
which are generally expressed as a boundary
condition for the propagators 
\cite{ZhangKurkijarviThuneberg,BuchholtzRainer,Kopnin,ROM}. 
The simplest surface model assumes a completely specular scattering of
the quasiparticles from the surface. 
This already leads to significantly improved results for the 
pinhole currents \cite{Viljas2}.
However, most surfaces are believed to
be microscopically rough, so that they scatter quasiparticles
diffusely. There are several model boundary conditions for such
surfaces, which all yield practically the same profiles 
$\Delta_\parallel(z)$ and $\Delta_\perp(z)$ 
\cite{ZhangKurkijarviThuneberg,BuchholtzRainer,Kopnin,ROM}.  
As in Ref.\ \onlinecite{Viljas2}, we 
use perhaps the simplest one, the ``randomly oriented mirror'' 
(ROM) model \cite{ROM}.
The gap-suppression effect also introduces spontaneous spin
currents which flow parallel to the surface \cite{ZhangKurkijarviThuneberg}. 
This is responsible for
the existence of a small spin-vector part in the mean-field 
self energy $\spin\Sigma^{\rm mf}_0=\nuvec_0\cdot\spin\sigmavec$, 
which depends on the Fermi-liquid parameters $F_l^a$ with odd $l$
\cite{Viljas2}.
The self-consistency procedure is simplest to do by using the
imaginary Matsubara energies, as briefly discussed in Appendix \ref{s.app1}.
In this part of the calculation, the inelastic relaxation rate
$\Gamma$ is taken to be infinitesimal.

Once the mean-field self-energies $\spin\Delta_{0}$ and 
$\spin\Sigma^{\rm mf}_{0}$ have been self-consistently 
computed (see Appendix \ref{s.app1}), the
coherence functions at real energies $\spin\gamma_0^{R,A}$ 
are calculated by integrating the Riccati equation 
\begin{equation} \label{e.mfriccati}
\begin{split}
\iu\hbar\vFvec\cdot\nablavec \spin{\gamma}^R_0
=&-2\epsilon^R\spin{\gamma}^R_0
-\spin{\Delta}_0 
-\spin{\gamma}^R_0\spin{\Delta}^\dagger_0\spin{\gamma}^R_0 \\
&+\spin{\Sigma}^{\rm mf}_0\spin{\gamma}^R_0
-\spin{\gamma}^R_0\spin{\td\Sigma}^{\rm mf}_0
\end{split}
\end{equation}
on several trajectories $\Rvec=u\kvechat$ passing through the pinhole
at $u=0$. To find the propagator at $u=0$, only integrations
from the bulk toward the wall are needed. 
In Eq.\ (\ref{e.riccati})
we again introduce the quasiparticle relaxation rate 
$\Gamma=2\Gamma_1/\hbar$ 
through $\epsilon^R=\epsilon+\iu(\hbar\Gamma/2)$.
We use the simple normal-state Fermi-liquid form
$\Gamma=a[(\pi\kB T)^2 + \epsilon^2]/(\tau_0\pi^2\kB^2)$, where
$\tau_0=1.14~\mu{\rm s}~{\rm mK}^2$ is obtained from viscosity measurements 
\cite{Wheatley75,Greywall86}, 
and $a$ is a free parameter of order unity.
The bulk solution of Eq.\ (\ref{e.mfriccati}) is easily found \cite{Eschrig}, 
and this is used as an initial condition.
For a junction with mirror-symmetry with respect to $z=0$, 
the $\lside$ and $\rside$ solutions at the junction satisfy 
$\spin\gamma_0^{R\rside}(-\kvechat,z=0)=-[\spin\gamma_0^{R\lside}(\kvechat,z=0)]^T$,
and thus the integration need only be done on one side.
The different spin-orbit rotations of the B phase 
on the two sides of the junction 
may be taken into account with the transformations
\begin{equation} \label{e.spinrot}
\begin{split}
\spin\gamma^{Ri}(\kvechat,z,\epsilon) &=
\spin U^i\spin\gamma_0^{Ri}(\kvechat,z,\epsilon)[\spin U^i]^T, \\
\spin\Delta^{i}(\kvechat,z) &=
\spin U^i\spin\Delta_0^{i}(\kvechat,z)[\spin U^i]^T, \\
\spin\Sigma^{{\rm mf},i}_{}(\kvechat,z) &=
\spin U^i\spin\Sigma_0^{{\rm mf},i}(\kvechat,z)[\spin U^i]^\dagger,
\end{split}
\end{equation}
where 
$\spin U^{i}$
$=\exp(-\iu\theta_L\nvechat^{i}\cdot\spin\sigmavec/2)$, and
$i=\lside,\rside$.
Finally, the functions $\spin\gamma^{Ri}(z=0)$ are inserted
into Eq.\ (\ref{e.ccomp}) to obtain the current amplitudes.

We note that the introduction of $\Gamma$ in Eq.\ (\ref{e.mfriccati})
is not self-consistent, and stress that our normal-state model
neglects all strong-coupling modifications on 
$\spin\Delta^{R,A}$ \cite{SereneRainer}.
This procedure should therefore be 
regarded only as a rough model for the description of the
quasiparticle-quasiparticle scattering.
However, the calculation still provides at least a semi-quantitative
model for studying the simultaneous effects of gap suppression and
inelastic processes on the pinhole currents.


\begin{figure}[!t]
\includegraphics[width=0.99\linewidth]{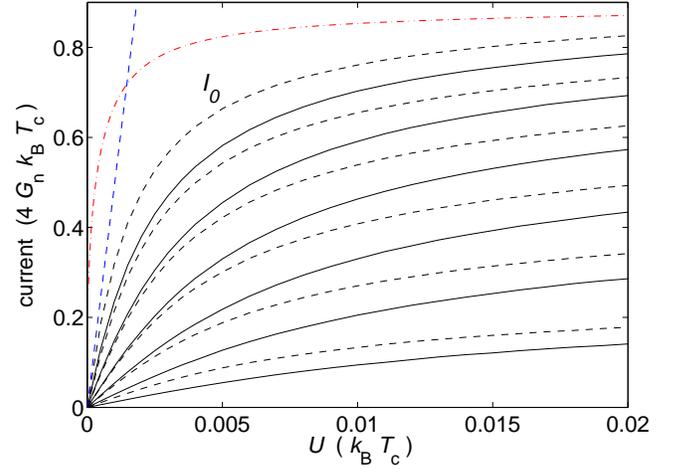}
\caption{The dc currents $I_0$ for temperatures
$T/\Tc=0.4,0.5,0.6,0.7,0.8,0.9$ in order of decreasing amplitude. 
The solid lines include the gap suppression effects, whereas in the
dashed lines it is neglected. The dash-dotted line is the
zero-temperature result of Eq.\ (\ref{e.zerot}), while
the straight dashed line corresponds to Eq.\ (\ref{e.arc}) at 
$T/\Tc=0.4$.
The results are similar for both
parallel and antiparallel $\nvechat^{\lside,\rside}$'s. 
Here $F_1^a=0$ and $a=1.6$.
}
\label{f.i0}
\end{figure}

\begin{figure}[!t]
\includegraphics[width=0.99\linewidth]{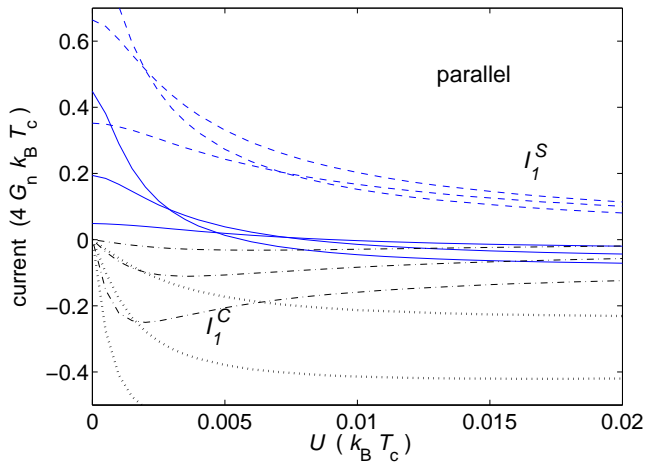}
\caption{The amplitudes $\IS_1$ and $\IC_1$ for 
$T/\Tc=0.4,0.6,0.8$ in order of decreasing amplitude in the
case of parallel $\nvechat^{\lside,\rside}$'s.
The solid and dashed lines are for $\IS_1$ with and without gap
suppression, respectively, and the dash-dotted and dotted lines are
the corresponding values for $\IC_1$.
Here $F_1^a=0$ and $a=1.6$.
}
\label{f.i1p}
\end{figure}

\begin{figure}[!t]
\includegraphics[width=0.99\linewidth]{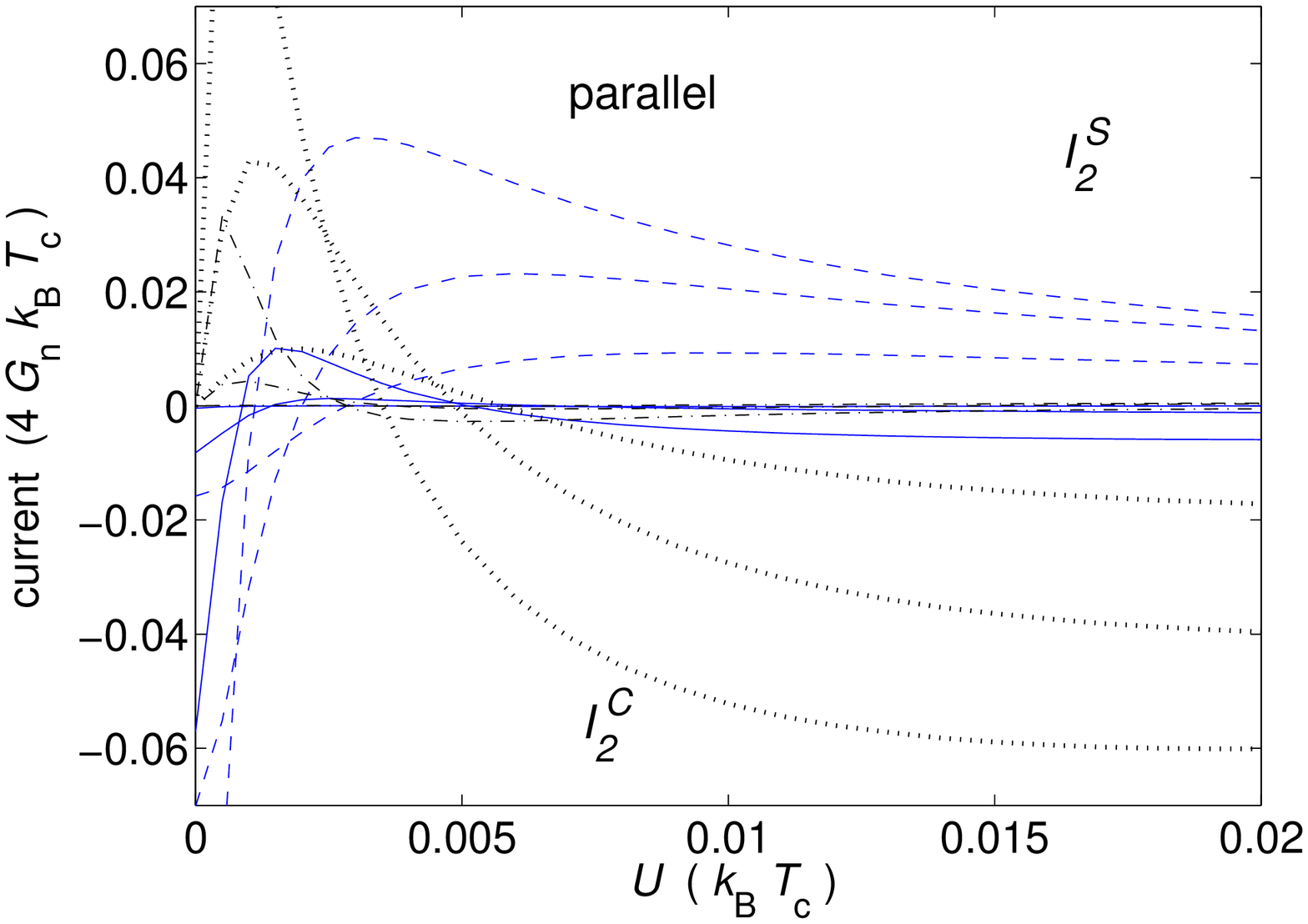}
\caption{Same as Fig.\ \ref{f.i1p} but for $\IS_2$ and $\IC_2$.}
\label{f.i2p}
\end{figure}

\begin{figure}[!t]
\includegraphics[width=0.99\linewidth]{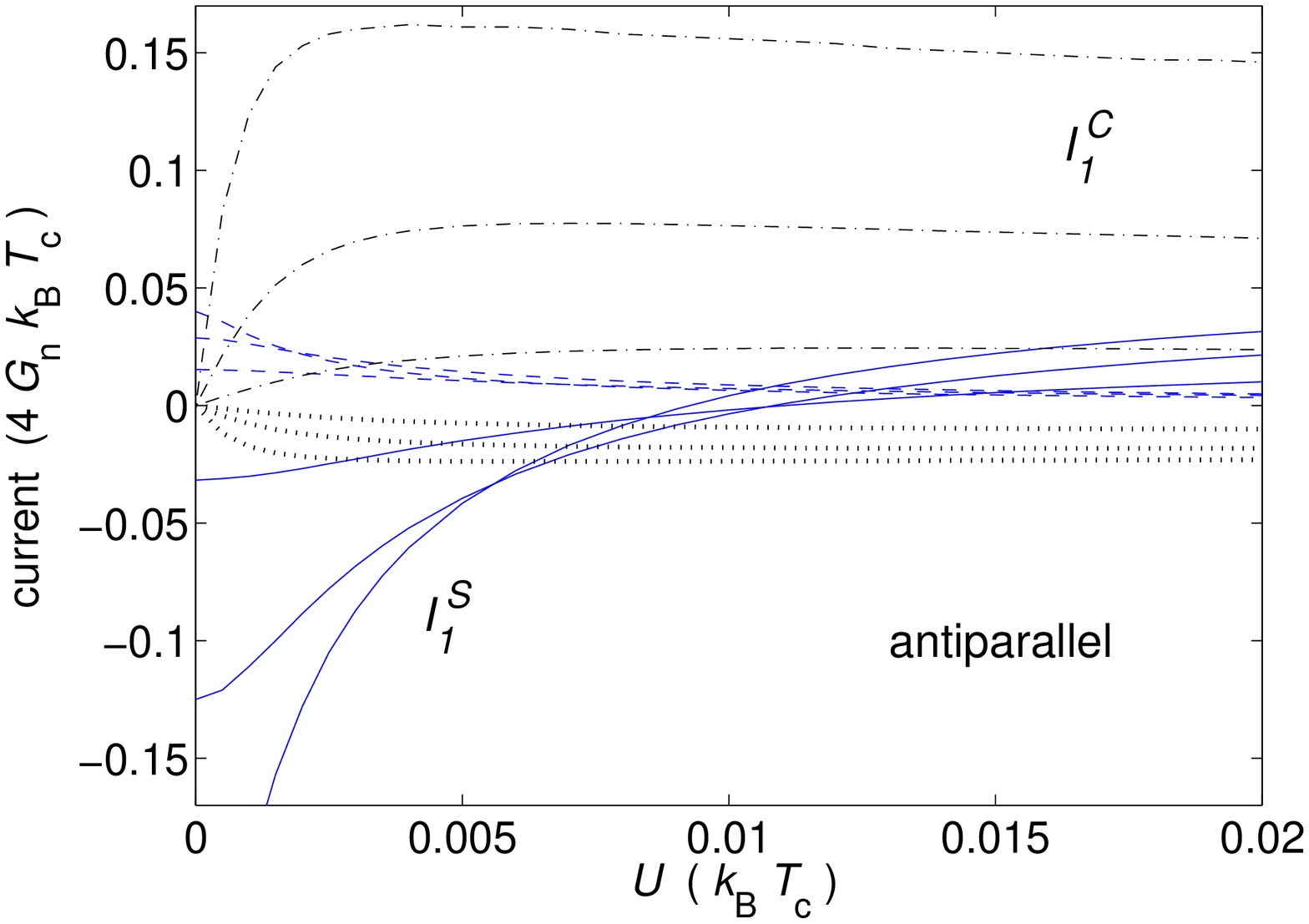}
\caption{Same as Fig.\ \ref{f.i1p} but for antiparallel $\nvechat^{\lside,\rside}$'s}
\label{f.i1ap}
\end{figure}

\begin{figure}[!t]
\includegraphics[width=0.99\linewidth]{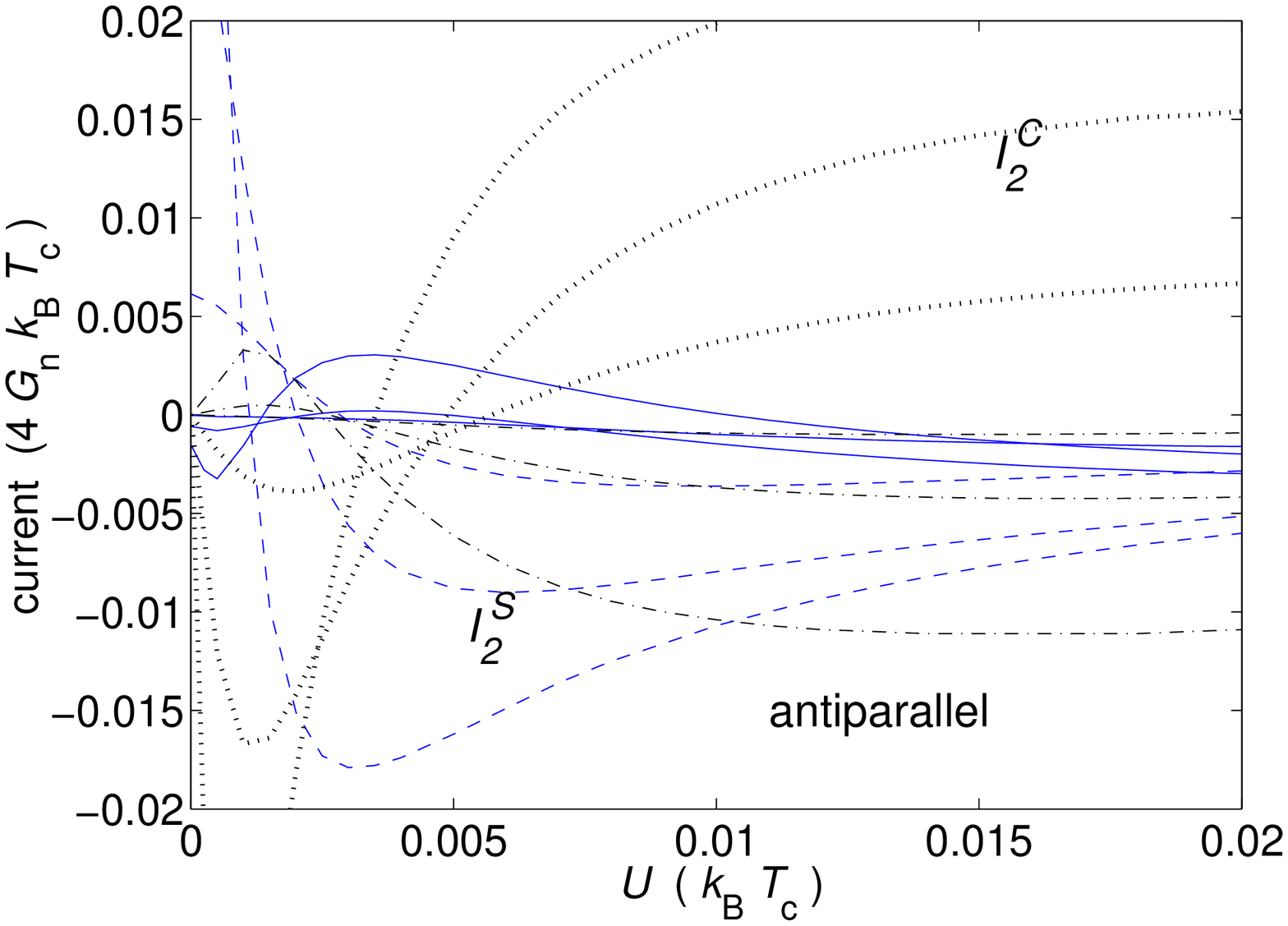}
\caption{Same as Fig.\ \ref{f.i2p} but for $\IS_2$ and $\IC_2$.}
\label{f.i2ap}
\end{figure}

\subsection{Current amplitudes}

Figures \ref{f.i0}-\ref{f.i2ap} show our results for the lowest-order 
current amplitudes $I_0$, $\IS_{1,2}$ and $\IC_{1,2}$ as a function of
the bias $U$. The amplitudes $\IS_m$ and $\IC_m$ 
for $m\geq 3$ may be set to zero since they
are negligibly small for all practical purposes. 
Figures \ref{f.i1p} and \ref{f.i2p} are for parallel 
$\nvechat$'s on the two sides of the contact
($\nvechat^\lside=\nvechat^\rside=\zvechat$) 
and Figs.\ \ref{f.i1ap} and \ref{f.i2ap} are for the antiparallel case
($-\nvechat^\lside=\nvechat^\rside=\zvechat$).
The $I_0$ amplitude shown in Fig.\ \ref{f.i0} is the same
for both cases. Actually, as our numerical calculation shows, 
the result for $I_0$ is practically independent
$R^{\lside,\rside}$. For the case where 
gap suppression is neglected, this independence 
was previously shown to be exact --- see
Eqs.\ (\ref{e.final}) and (\ref{e.dch}).

In the calculations, we have always assumed 
$F_l^a=0$ for $l\geq 3$, while the effect of the remaining 
$F_1^a$ with values between $-1\ldots0$ on the pinhole currents is 
at most a few percent of their total
amplitude \cite{Viljas2} --- in the figures
of this paper $F_1^a=0$ also. 
The quasiparticle relaxation parameter is chosen as $a=1.6$, since
this is the value which gives the best fit to the relevant experiments ---
see Sec.\ \ref{s.comparison}.
Although the behavior of the current amplitudes 
for $U\lesssim\hbar\Gamma$ depends
strongly on the choice of $a$, the asymptotic behavior for 
$U\gg\hbar\Gamma$ does not.

We have only calculated the currents numerically down to $T/\Tc=0.4$, 
although Fig.\ \ref{f.i0} shows the additional zero-temperature
result of Eq.\ (\ref{e.zerot}). This is because the required number of 
terms in the MAR sum [Eq.\ (\ref{e.ccomp})] is proportional to
$\Delta/\Gamma_0$, which is on the order of thousands already 
at $T/\Tc=0.4$. 
We have taken into account all terms up to $n=4\Delta/\hbar\Gamma_0$. 
Actually, this is only necessary for 
$U\ll\hbar\Gamma$ where the number of successive 
Andreev reflections is limited by
relaxation. In this regime it would also be possible to use 
approximate schemes instead of the full expression (\ref{e.ccomp}).
On the other hand, for $U\gtrsim\hbar\Gamma$ a maximum 
of $\sim \Delta/U$ terms should 
be enough. The energy cutoff for the coherence functions 
was typically chosen at around $10\kB\Tc$, and the density of
energy discretization points was highest close to the gap edges, where
the highest accuracy is needed.
For energies in between these points, linear interpolation was used 
in computing the current amplitudes with Eq.\ (\ref{e.ccomp}).
The number of Gaussian polar angles used in angular integrals was usually
eight.

In the limit $U\rightarrow 0$ the cosine amplitudes $\IC_m$ all
vanish (linearly in $U/\hbar\Gamma_0$) as they should, since
the equilibrium current  
$I_{\rm s}(\phi)$ must satisfy the time-reversal symmetry 
$I_{\rm s}(-\phi)=-I_{\rm s}(\phi)$.
Upon inserting the $\IS_m$ amplitudes to Eq.\ (\ref{e.fullc}) in this limit,
the current-phase relations $I_{\rm s}(\phi)$ of Ref.\ \onlinecite{Viljas2}
are quite accurately reproduced, 
when we make the replacement $\omegaJ t\rightarrow \phi$. 
Note, in particular, that in the antiparallel case with no gap
suppression, the 
second-order amplitudes are of equal magnitude with the first-order
ones. Thus $I_{\rm s}(\phi)$ exhibits a strong ``$\pi$ state'' \cite{Yip}.
When the gap-suppression effect is taken into account, the $m=2$ 
amplitudes tend to be very strongly suppressed, and in this case
the ``$\pi$ state'' only presents itself at very low temperatures
\cite{Viljas2}.
Thus as a first approximation, 
$\IS_2$ may often be neglected in comparison with $\IS_1$.
However, apart from the limit $U\ll \hbar\Gamma$ of very low biases, 
the cosine amplitude $\IC_1$ as well as the dc component $I_0$ are equally 
large as the sine amplitude $\IS_1$. 
Therefore the models 
which are based on $\IS_1$ alone --- like that of Ref.\
\onlinecite{Viljas2} --- are only valid for $U\ll\hbar\Gamma$
and $T$ close to $\Tc$.


\begin{figure}[!t]
\includegraphics[width=0.99\linewidth]{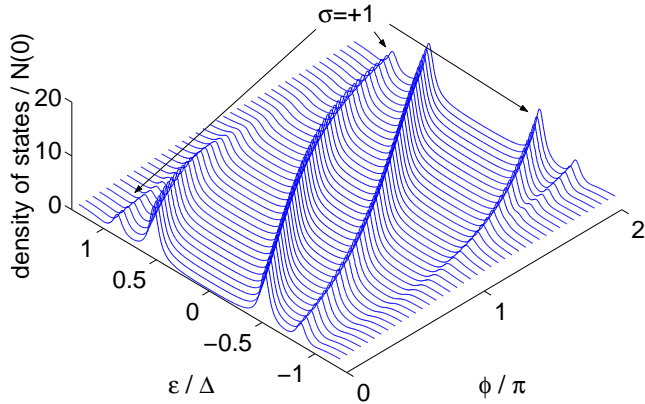}
\caption{
Local density of states inside the pinhole 
for $\hat k_z=0.93$ and $-\nvechat^l=\nvechat^r=\zvechat$
when gap suppression at a diffusive wall is included. 
Here a large, arbitrary value $a\approx 75$ was chosen for purposes of
illustration, in order to make the 
width $\sim\hbar\Gamma$ of the bound-state peaks better observable.
Among the peaks, those corresponding to the $\sigma=+1$ ($\sigma=-1$) 
spin branch are the ones apparently shifted toward smaller (larger) $\phi$.
Note that the slope of the bound states is not so steep as for Eq.\
(\ref{e.states}). Also, for given phase difference and spin band, more
than one bound state can exist simultaneously.
}\label{f.didos}
\end{figure}

\subsection{Bound states} \label{s.bstates}

In Sec.\  \ref{s.nosupp} we saw that in the adiabatic regime
$U\ll\Delta$, the currents in the point contact may simply be described
in terms of the equilibrium bound states. 
It should be possible to generalize this description to the case where
gap-suppression is included. 
To study this, we have calculated numerically the density of
quasiparticle states at the pinhole.
The bound states
are given by the sub-gap peaks in the $\kvechat$-resolved
local density of states (or spectral density)
\begin{equation} \label{e.dos}
N(\kvechat,\Rvec;\epsilon)=-\frac{N(0)}{\pi}\im{\Tr}
\spin g_{}^R(\kvechat,\Rvec;\epsilon).
\end{equation}
These peaks correspond to poles of $\nam g^R$, which depend on the
phase difference $\phi=\varphi^\rside-\varphi^\lside$
and the rotation matrices through
$\psi_{ij}=R_{\mu i}^{\lside}R_{\mu j}^{\rside}$.
The width of the peaks is on the order of the relaxation rate
$\hbar\Gamma$.
Although the spectral weight of the bound states is spread 
around the pinhole up to distances of order 
$\hbar\vF/\sqrt{\Delta^2-\epsilon^2}$,
we only calculate Eq.\ (\ref{e.dos}) inside the hole, at $\Rvec=\zerovec$.

Figure 
\ref{f.didos} illustrates the results for one choice of $\psi_{ij}$ and 
$\kvechat$ with $\phi=0\ldots2\pi$
when the gap suppression effect of a diffusive surface is included. 
Since $\psi_{ij}\neq\delta_{ij}$, the bound-state peaks in the 
densities show clearly a
spin-splitting between the ``$\sigma=\pm 1$'' condensates, 
as in the simple analytic result (\ref{e.states}). 
Since the splitting angle depends on $\kvechat$, an average
of $N(\kvechat,\epsilon)$ over the Fermi surface in fact 
leads to the a formation of a wide \emph{band} of bound state energies.
The bound states for $-\kvechat$ are obtained by using the 
time-reversal symmetry
$\kvechat\rightarrow-\kvechat$,
$\phi\rightarrow-\phi$,
$\sigma\rightarrow-\sigma$, 
and the ``particle-hole'' symmetry
$\kvechat\rightarrow-\kvechat$, $\epsilon\rightarrow-\epsilon$,
which follows from the symmetries of the equations of motion and 
the geometry \cite{ZhangKurkijarviThuneberg}.

Compared to Eq.\ (\ref{e.states}), it is seen that
the gap suppression modifies the bound states 
such that they are always at energies $|\epsilon|<\Delta$. 
In the bulk 
$N(\kvechat,\epsilon)\approx
2N(0)\theta(|\epsilon|-\Delta)|\epsilon|/\sqrt{\epsilon^2-\Delta^2}$ has 
divergences at $|\epsilon|=\Delta$, but in the middle of the junction 
most of the spectral weight is now in the bound states even at $\phi=0$.
Thus there are several branches of bound states coexisting 
simultaneously for given $\phi$. Accordingly,
Eq.\ (\ref{e.statecurrent}) for the current should be modified by
replacing 
$\epsilon_{\kvechat,\sigma}(\phi)$ with
$\epsilon_{q,\kvechat,\sigma}(\phi)$, and by 
adding a sum over the branch index $q$.
Since the energies $\epsilon_{q,\kvechat,\sigma}(\phi)$ for a given branch
in the range $(-\Delta,\Delta)$ are now mapped to phase 
differences in the range $(-\infty,\infty)$,
their slopes are not so steep. Therefore, one would expect that 
the dc currents are generally smaller when the gap suppression 
effect is taken
into account. As seen in Fig.\ \ref{f.i0}, this is usually the case.

\section{Comparison to experiment} \label{s.comparison}

In this section we present a brief comparison of the above theory
to available experimental data on the current-pressure characteristics
in $^3$He-B weak links \cite{Steinhauer,Simmonds}. 
There are a couple of basic things to note about the experiments. 
First, as already mentioned, it is 
difficult to manufacture apertures which would 
satisfy the requirements of a pinhole very well \cite{DavisPackardRMP}.
The apertures with $d\approx 100$ nm in a 50 nm membrane used in 
Ref.\ \onlinecite{Simmonds} are rather close, at least compared with the
0.25 $\mu$m wide slits in a 0.1 $\mu$m membrane 
of Ref.\ \cite{Simmonds}. 
Second, in order to ensure leak-proofness, 
the pressure biases are limited to
very low values where $U\ll\Delta$. 
The experiments of Ref.\ \onlinecite{Simmonds} are even
restricted to $U\ll\hbar\Gamma\ll\Delta$.
Therefore, the ``excess current''
limit of Eq.\ (\ref{e.exc}), for example, seems not practically 
achievable in superfluid $^3$He.
Nevertheless, the data of Refs.\ \onlinecite{Steinhauer,Simmonds} are
enough to make a comparison between the most important features of the
theoretical and experimental current-pressure characteristics.

\begin{figure}[!t]
\includegraphics[width=0.99\linewidth]{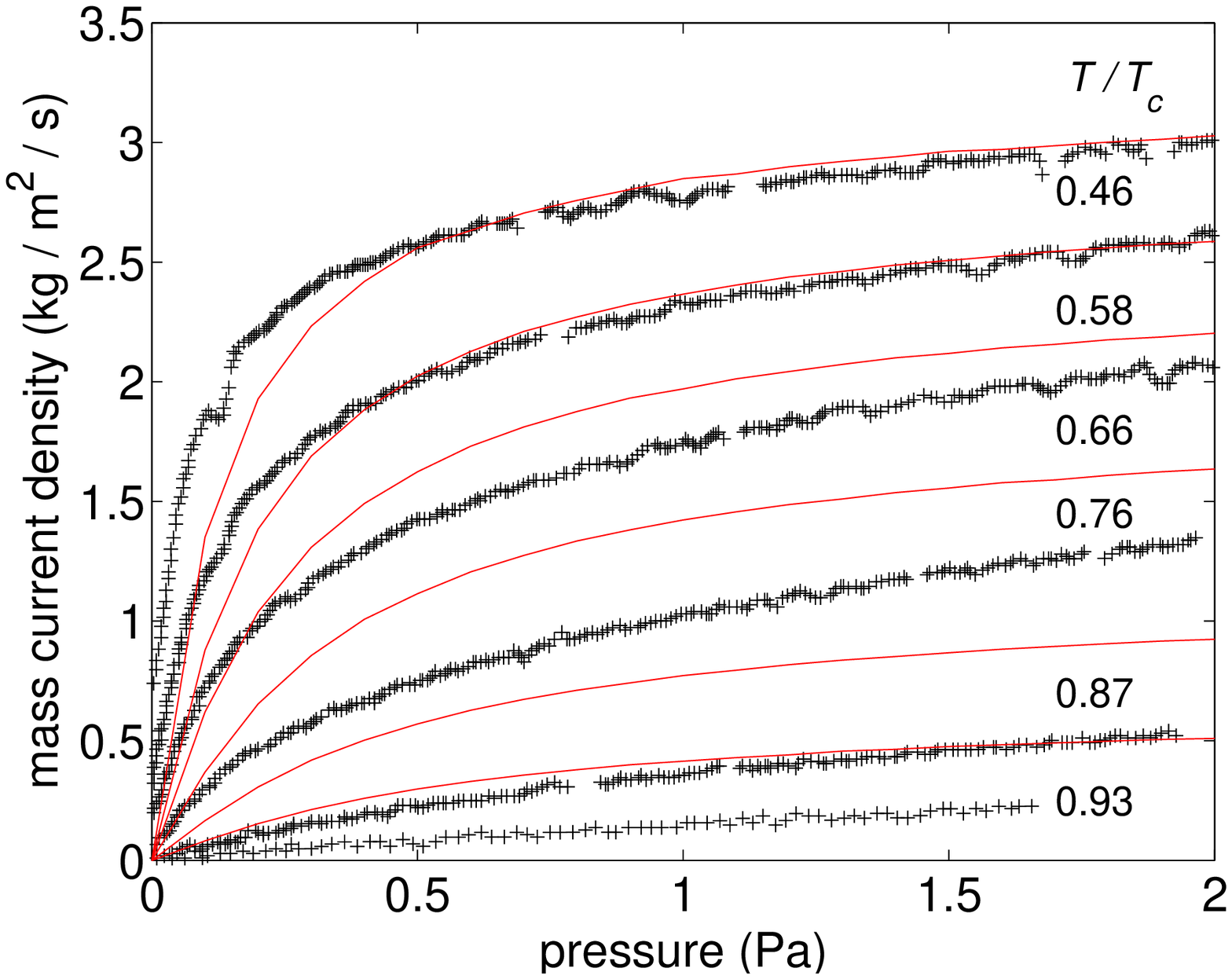}
\caption{
Comparison of the data (+ signs) from Ref.\ \onlinecite{Steinhauer}
to the pinhole theory (solid lines) for a diffusive surface 
and $a\approx0.27$ at indicated temperatures.
}\label{f.ipfits}
\end{figure}

Figure \ref{f.ipfits} shows a comparison between the data of 
Ref.\ \onlinecite{Steinhauer} and the numerical pinhole calculation of
$I_0(U)$.
A diffusive surface is assumed, and we use the parameter $a\approx0.27$.
Again we note that although the slope at $U=0$
depends strongly on $a$, the asymptotic behavior 
for $U\gg\hbar\Gamma$ does not.
Indeed, at $U\gtrsim\hbar\Gamma$ the agreement
is rather good for any value of $a$ on the order of unity, 
although a perfect fit 
for all temperatures simultaneously cannot be achieved. 
Note also that the experimental currents are not even approaching 
zero in the limit $U\rightarrow 0$.
However, a better fit can hardly be expected, since the apertures
used in these experiments were far from good pinholes.
In fact, in sufficiently large apertures one would expect a 
transition into a regime
where the dissipation is best described in terms of \emph{phase slips} by
vortices rather than MAR. Studying this transition would
be interesting, but computationally very demanding.
In any case, the overall form of $I_0(U)$ is very similar to the
experimental results, and the order-of-magnitude
agreement on both axes is surprisingly good.
This gives strong support to the expectation 
that the dominating dissipation mechanism 
in small apertures of superfluid $^3$He is the MAR process,
analogously to superconductor point contacts.

\begin{figure}[!t]
\includegraphics[width=0.99\linewidth]{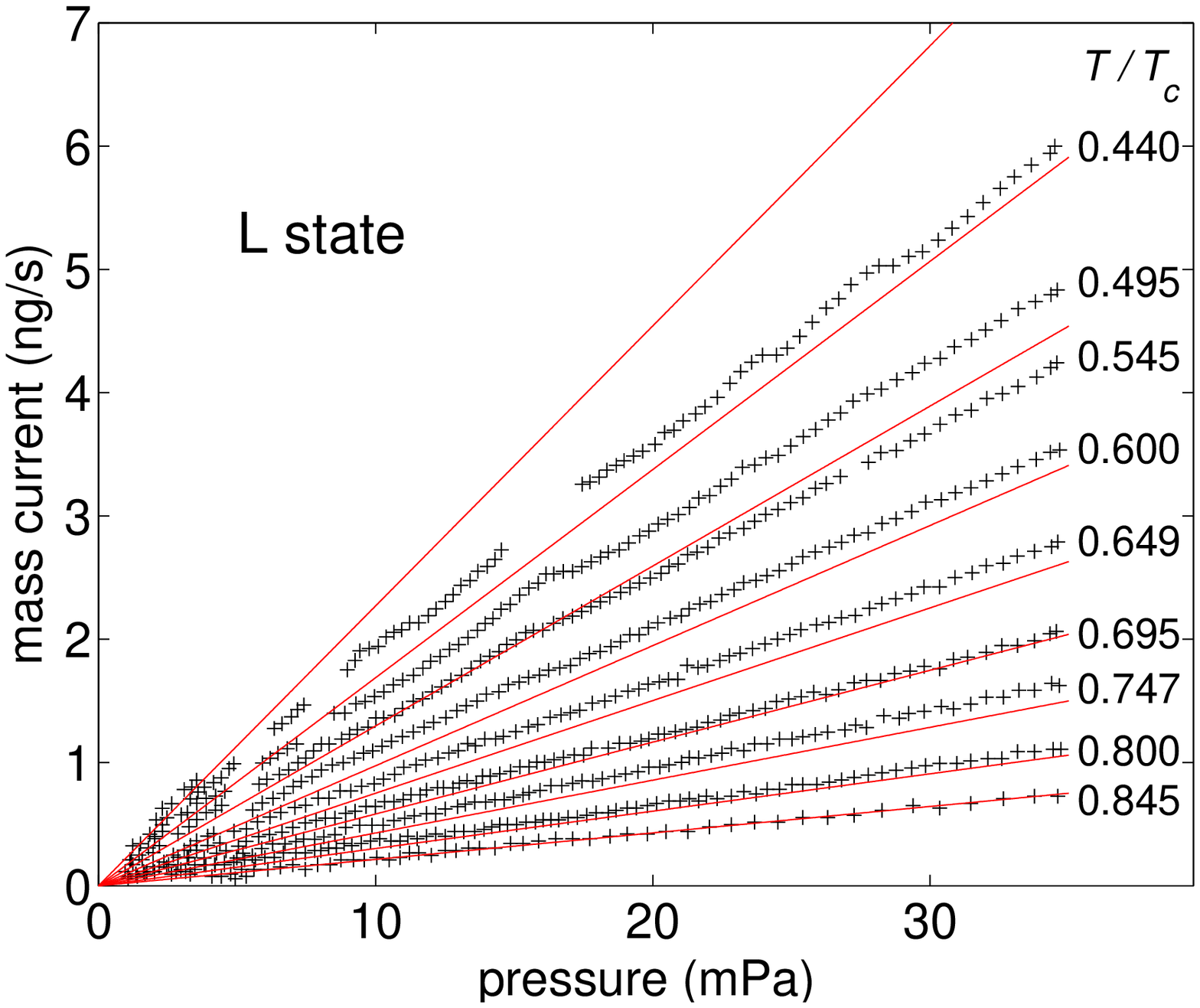}
\caption{
Comparison of the L state data (+ signs) from Ref.\ \onlinecite{Simmonds}
to the pinhole theory (solid lines) for a diffusive surface and
$a\approx1.6$ at indicated temperatures.
}\label{f.ipfitl}
\end{figure}

The apertures used in the experiments reported in Ref.\ \onlinecite{Simmonds}
are somewhat better approximations to pinholes.
Therefore we have chosen to use these data to estimate the 
value of $a$ for the numerical calculations of the previous Section.
The experiments were carried out in the limit $U\ll\hbar\Gamma$, and 
according to Eq.\ (\ref{e.arc}), the current should be linear in the
bias. For the ``L state'' this is rather well satisfied, and 
a fit to the L-state data gives $a\approx 1.6$, when gap suppression at a 
diffusive surface is taken
into account \cite{Viljas5}.
The fit is shown in Fig.\ \ref{f.ipfitl}.
The increasing low-temperature deviations in this fit 
may be partly due to the insufficiency of our
normal-state model for $\Gamma$.
The currents for the ``H state'' reported in Ref.\
\onlinecite{Simmonds} are larger and more nonlinear than in the 
L state. We suspect that part of this 
nonlinearity may be due to the additional
dissipation effects described by Eq.\ (\ref{e.avg}) \cite{Viljas5}.
Another possibility is that the experimental 
apertures already deviate so strongly from
pinholes, that the dependence of the bound states on textures
is not sufficiently described by simple phase shifts.

\section{Conclusions and discussion} \label{s.disc}

In conclusion, we have presented an analysis of pressure-biased weak
links between two volumes of superfluid $^3$He-B by using the pinhole
model of a short, ballistic point contact. We showed how the $s$-wave
results of Refs.\ \onlinecite{GunsenheimerZaikin,AverinBardas} may easily be
reproduced and generalized to the $p$-wave case by
parametrizing the Green functions with the so-called coherence functions. 
In the case where the gap
suppression at surfaces is neglected, we calculated the current
amplitudes analytically in several limits. In the general case, the
order parameters and the pinhole currents were calculated numerically.
Comparison to experiments gives strong support to the existence
of the MAR effect in pressure-biased weak links of superfluid 
$^3$He \cite{Steinhauer}.
We also predict the existence of additional dc current contributions,
which result from the excitation of collective order-parameter modes.
These ``anisotextural'' effects are discussed elsewhere in more 
detail \cite{Viljas5}.

In order to improve upon the results of the present paper, one should
take into account the strong-coupling effects in a more detailed
way than with our ``normal-state'' model for the quasiparticle relaxation
rate. One should also consider apertures of finite size, at least by 
computing the bound-state spectra in equilibrium to see if textures have some
significant effect on them in this case. A dynamical
calculation for the finite-size aperture should also be performed, but
this already seems to approach the limits of practical feasibility.
A fully self-consistent calculation of the anisotextural effects in a
pressure-biased pinhole array also appears to be very difficult.
However, until such improvements are made, a parameter-free
comparison with the experimental data cannot be expected. 

With somewhat less effort, the pinhole heat conductivity \cite{EskaHeat} 
or spin currents could be studied by starting from 
Eq.\ (\ref{e.intermediate}). 
The current-noise 
properties \cite{AverinImam} of a $^3$He pinhole
could also be of some interest, for
example in the design of accurate superfluid $^3$He gyrometers 
\cite{AvenelMukharsky}.


\acknowledgments

Prof. E. V. Thuneberg is gratefully acknowledged for fruitful
discussions concerning Secs.\ \ref{s.nosupp} and \ref{s.bstates} 
--- especially for pointing
out the result of Eq.\ (\ref{e.zerot}).
The Center for Scientific Computation (CSC) is thanked
for computing resources.

\begin{appendix}

\section{Equilibrium equations} \label{s.app1}

Close to the planar wall at $z=0$ 
the mean-field self energies and the coherence functions must be iterated 
self-consistently.
Since we are concerned with an equilibrium system, it is easiest to
do this by using the Matsubara technique, where one makes an
analytical continuation from the real energies
$\epsilon^{R,A}=\epsilon\pm\iu 0^+$ to the 
imaginary Matsubara energies $\pm\iu|\epsilon_m|$, where
$\epsilon_m(T)=\pi\kB T(2m+1)$ and $m=0,\pm1,\pm 2,\ldots$
\cite{SereneRainer}.
The coherence functions are then obtained from
Eq.\ (\ref{e.mfriccati}), where $\epsilon^R\rightarrow\iu\epsilon_m$
and $\spin\gamma_0^R(\epsilon)\rightarrow\spin\gamma_0(\epsilon_m)$.
(Here we take $\Gamma_1$ as infinitesimal.)
The self-energies are of the form
$\spin\Delta_0=\Deltavec_0\cdot\spin\sigmavec\iu\spin\sigma_2$ and
$\spin\Sigma^{\rm mf}_0=\nuvec_0\cdot\spin\sigmavec$.
If we use the symmetry $\td\nuvec_0=-\nuvec_0$ 
\cite{SereneRainer,Viljas2}, the solution of Eq.\ (\ref{e.mfriccati}) 
may be parametrized by 
$\spin\gamma_0=\gammavec_0\cdot\spin\sigmavec\iu\spin\sigma_y$, and
the equation becomes
\begin{equation} \label{e.matsuriccati}
\begin{split}
\iu\hbar\vFvec\cdot\nablavec\gammavec_0 =& -2\iu\epsilon_m\gammavec_0
-\Deltavec_0-2(\gammavec_0\cdot\Deltavec_0^*)\gammavec_0 \\
&+(\gammavec_0\cdot\gammavec_0)\Deltavec_0^*+2\iu\nuvec_0\times\gammavec_0.
\end{split}
\end{equation}
Due to translational symmetry along the wall, the solution and the
self-energies only depend on the coordinate $z$. 
The self-consistency equations
are given by \cite{SereneRainer}
\begin{equation} \label{e.deltaeq}
\begin{split}
\Deltavec_0(\kvechat,z)&=3\kB T V_1
\sum_{|\epsilon_m|<\epsilon_c}\left\langle(\kvechat\cdot\kvechat') 
\fvec_0(\kvechat',z,\epsilon_m)\right\rangle_{\kvechat'} \\
\end{split}
\end{equation}
where the $p$-wave pairing interaction 
$V(\kvechat\cdot\kvechat)=3V_1\kvechat\cdot\kvechat'$ was assumed, and
\begin{equation} \label{e.nueq}
\begin{split}
\nuvec_0(\kvechat,z)&=3\kB T
\sum_{|\epsilon_m|<\epsilon_c}\left\langle A_a(\kvechat\cdot\kvechat') 
\gvec_0(\kvechat',z,\epsilon_m)\right\rangle_{\kvechat'}.
\end{split}
\end{equation}
In these 
$\langle\cdots\rangle_{\kvechat}=\int(\upd\Omega_\kvechat/4\pi)(\cdots)$,
$\gvec_0={\Tr}_2(\spin\sigmavec\spin g_0)/2$,
$\fvec_0=-{\Tr}_2(\spin\sigmavec\spin g_0\iu\spin\sigma_2)/2$
where
the
upper diagonal and off-diagonal Nambu components of the 
Matsubara propagator  $\nam g(\kvechat,z,\epsilon_m)$ are given by
$\spin g_0=-\pi\iu(\spin 1+\spin\gamma_0\spin{\td\gamma}_0)(\spin
1-\spin\gamma_0\spin{\td\gamma}_0)^{-1}$ and
$\spin f_0=-2\pi\iu\spin\gamma_0(\spin 1-\spin\gamma_0\spin{\td\gamma}_0)^{-1}$.
The conjugation symbol ``$~\td~~$'' now means 
$\spin{\td\gamma}_0(\kvechat,\epsilon_m)=\spin\gamma_0(-\kvechat,\epsilon_m)^*$
$=[\spin\gamma_0(\kvechat,-\epsilon_m)]^\dagger$.
The coupling constant $V_1$ may be eliminated in the usual way by
noting that at
$T\rightarrow\Tc^-$ the gap vector $\Deltavec_0$ is small and 
$\fvec_0\approx\pi\Deltavec_0/|\epsilon_m|$. Thus $\Deltavec_0$ may be
canceled and one finds
\begin{equation} \label{e.elim}
\begin{split}
V_1^{-1}
&\approx\pi\kB T_c\sum_{|\epsilon_m|<\epsilon_c}|\epsilon_m|^{-1} + \ln(T/\Tc)
\end{split}
\end{equation}
where $\epsilon_m=\epsilon_m(T)$.
Inserting this into Eq.\ (\ref{e.deltaeq}), the cutoff $\epsilon_c$
may be taken to infinity --- see Eq.\ (12) in 
Ref.\ \onlinecite{Viljas2}.
The function
$A_a(x)=\sum_{l=0}^{\infty}F_l^a[1+F_l^a/(2l+1)]^{-1}P_l(x)$, where
$P_l(x)$ are the Legendre polynomials.
From symmetries of the planar wall geometry it follows that if 
$\kvechat$ is in the $xz$ plane, so are $\fvec_0$, 
$\Deltavec_0(\kvechat)=\Delta_\parallel\hat
k_x\xvechat+\Delta_\perp\hat k_z\zvechat$ and $\gammavec_0$,
while $\gvec_0$ and 
$\nuvec_0(\kvechat)=\nu_{0y}(\kvechat)\zvechat$ are perpendicular
to it.
Also due to symmetries, all $F_l^a$ with even $l$ drop out of the
theory, and for $l\geq 3$ we assume them to be zero \cite{Viljas2}.

In the bulk $\Deltavec_0$ is constant and $\nuvec_0=0$, so that 
Eq.\ (\ref{e.matsuriccati}) is easily solved,
and this solution is used
as an initial condition on trajectories starting from the bulk. 
On the wall side, one needs a
boundary condition to compute the ``outgoing'' coherence functions 
from the ``incoming'' functions. We use the ROM boundary condition, 
which is explained in Ref.\ \onlinecite{ROM}. 
This is easiest to express in terms of the
components of the propagator $\nam g(\kvechat,z,\epsilon_m)$, 
rather than directly with the
coherence functions $\gamma_0(\kvechat,z,\epsilon_m)$. The relation 
$\spin\gamma_0=-(\iu\pi-\spin g_{0})^{-1}\spin f_0$ is then useful for
computing the initial condition for $\spin\gamma_0$ on the outgoing
trajectories \cite{VorontsovSauls}.
A specularly reflecting surface is much simpler to implement, since it
only leads to the the continuity condition
$\spin \gamma_0(\ul{\kvechat})=\spin\gamma_0(\kvechat)$, 
where $\ul{\kvechat}=\kvechat-2(\kvechat\cdot\zvechat)\zvechat$
is the specularly reflected direction.
As shown in Ref.\ \onlinecite{Viljas2}, the pinhole current amplitudes
for specular and diffusive surfaces usually have only minor (at
most a few percent) differences.
Thus, in practice the implementation of a diffusive surface 
may not be worth the trouble.




\section{Folding products} \label{s.app2}

The quasiclassical folding product between objects of the type
$\nam A(\epsilon,t)$ is defined in Ref.\ \onlinecite{SereneRainer}.
This result is generalized to multiple products of $n$ objects
$\nam A_j(\epsilon,t)$, $j=1,2,\ldots,n$ as follows
\begin{equation} \label{e.mixfold}
\begin{split}
(\nam A_1&\fold\cdots\fold \nam A_n)(\epsilon,t) \\
&=\prod_{j=1}^n \nam A_j[\epsilon +\frac{\hbar}{2\iu}
(\cdots+\partial_{j-1}-\partial_{j+1}-\cdots), t_j],
\end{split}
\end{equation}
assuming all the Fourier-transformations leading to the ``mixed''
representations $\nam A_j(\epsilon,t)$ exist.
Here $\partial_j$ refers to a derivative with respect to the $j$th
time variable $t_j$.
When the time dependence of $\nam A_j(\epsilon,t)$ is harmonic
[\emph{i.e.}, $\propto\exp(\iu\epsilon t/\hbar)$],  
Eq.\ (\ref{e.mixfold}) yields the folding product analytically.
In equilibrium all time derivatives vanish, and the folding product 
becomes a simple matrix product.



\end{appendix}

\end{document}